Direction Dependence and Diurnal Modulation In Dark Matter Detectors

Richard J. Creswick[1], Shmuel Nussinov[2] and Frank T. Avignone III[1]

[1]Department of Physics and Astronomy
University of South Carolina
Columbia, SC 29208, USA

[2]Tel Aviv University, Sackler School Faculty of Sciences
Ramat Aviv, Tel Aviv 69978, Israel and Tel Aviv University, Israel and
Schmid College of Science, Chapman University, Orange, California 92866, USA

**Abstract**
In this paper we study the effect of the channeling of ions recoiling from collisions with weakly interacting massive particles (WIMPs) in single crystal detectors. In particular we investigate the possibility that channeling may give rise to diurnal modulations of the counting rate as the Earth rotates relative to the direction of the WIMP wind, and the effect that channeling has on the "quenching factor" of a detector.

## I. Introduction

Astrophysical evidence strongly suggests that most of the matter in the universe is non-baryonic "dark matter". Direct searches for dark matter particles in low-background underground detectors has been ongoing for several decades. [1-3] Detector nuclei struck by dark matter 'WIMPs' (Weakly Interacting Massive Particles) with virial velocities recoil with low energies (~ 1-100 keV) depending on the WIMP and nuclear masses. The resulting energy deposition is *local* and occurs uniformly over the detector material. Even in cases in which extreme care is taken to reduce the backgrounds due to cosmic muons, spalation neutrons and natural radioactivity, low energy signals alone cannot provide a clear WIMP signature. However, the motion of the detector relative to the galactic center can provide a time-dependent modulation of low-energy events characteristic of WIMPs.

An annual modulation of the WIMP flux due to the motion of the Earth around the Sun was suggested some time ago in a seminal paper by Drukier, Freese and Spergel [4]. They assumed a Gaussian WIMP velocity distribution, isotropic in the frame of the galaxy, with a width comparable to the local velocity of the solar system, about 240 km/s. When the motion of the Sun through the galactic halo, is compounded with the motion of the Earth around the Sun, the flux of WIMPS varies annually by about ±7%. Such a modulation has been reported by the DAMA-LIBRA Collaboration [5], but not in other WIMP searches [6-11]. Various particle physics models including the "exciting WIMP model", light WIMPs with masses of order 10 GeV, and spin dependent WIMP-nuclei interactions, attempt to reconcile the bounds from these other experiments with DAMA-libra [12-15].

In addition to the modulation in the WIMP flux due to the orbital motion of the Earth, the anisotropic angular distribution of WIMP momenta as observed from a detector on the Earth might also provide a unique WIMP signature. All current large-scale experiments measure recoil energies only.

Sekiya et al., [16-18] have described the development of a directional dark matter search using low symmetry (monoclinic) stilbene crystals to look for daily modulation induced by channeling. In a neutron scattering experiment they found about a 7% variation in the detector's response as a function of the recoiling ion's direction for recoil energies in the range 4-6 keV. In a recent paper [19] we discussed the diurnal modulation effect based on the directionality of the WIMPs and channeling of the recoil ions.

When an ion channels between two atomic planes in a crystal it deposits more of its energy in electromagnetic modes (e.g. scintillations or ionization) and less in the form of phonons. The quenching factor, $Q$, is the ratio of the electromagnetic energy deposited by the ion to that deposited by a photon of the same energy. When an ion channels, Q can approach unity, whereas for an unchanneled ion $Q$ is much smaller. Channeling has been invoked in order to explain the discrepancy between the DAMA-LIBRA results and other WIMP searches.

Several effects mitigate against a strong correlation between the distribution of recoils and the direction of the WIMP wind. First, the velocity **v** of the WIMP is the vector sum of the fixed wind **w** and a random virial velocity, **u**, of roughly equal magnitude. Hence at any given time the direction of the WIMP impinging upon the crystal has a broad distribution that is only peaked around **w**. The directionality effect is further diluted by the recoil of the elastically scattered nucleus, which we assume is isotropic in the CM frame. Finally, the directionality effect is further reduced by the nuclear form factor, relevant for heavier nuclei, which suppresses scattering a high momentum transfer.

These purely kinematic issues are encountered in *any* attempt to
utilize WIMP anisotropy and are discussed next in Section II below. Section III addresses the effects of channeling. We adopt a simple model for channeling that incorporates some of the basic physics; for a more detailed study the reader is referred to the recent work of Bozorgnia, Gelmini, and Gondolo [20].

Upon folding the expected yield as a function of recoil direction with the distribution of recoils obtained in the previous section we find in general diurnal modulations far smaller than our original optimistic expectations [16]. This accords with the preliminary detailed estimates of channeling effects in general and daily modulations in particular. The basic reason underlying this is the high degree of symmetry of the lattices considered and the simple cubic lattice of Na-I in particular. This together with the wide distribution of the directions of the recoil ions makes for some channeling along the many crystal planes (100, 110 etc.) at any given time. Hence the daily variations are minimized. However, we find that in certain models with "light" WIMPs the requirement imposed by the limited sensitivity of the experiment that recoil energies must exceed some minimal threshold values can strongly align all the relevant directions and thereby largely restore the effect. We comment on these issues and on the possible motivation for light and in particular asymmetric WIMP models in the last section IV.

**II. Recoil Distribution**

We assume, following the standard halo model, that the velocity distribution of WIMPs in the rest frame of the Earth is essentially a Gaussian,

$$f(\mathbf{v}) = \frac{1}{\left(\pi u_0^2\right)^{3/2}} \exp\left[-\frac{|\mathbf{v}-\mathbf{w}|^2}{u_0^2}\right] \qquad (1)$$

where **w** is the WIMP wind velocity, (-**w** is the velocity of the Earth relative to the galactic center), $|\mathbf{w}| \sim 240$ km/s and $u_0 \sim |\mathbf{w}|$ is the spread in the distribution. In order to simplify the analysis we ignore the cutoff in the random component of the velocity at the galactic escape velocity; this has little impact on our results.

Since the velocity distribution is independent of the mass of the WIMP, the energy of the WIMP scales with the WIMP mass, $m_X$, and is given by

$$D(E/m_X) = \frac{1}{\sqrt{\pi} u_0 w} \left\{ \exp\left[-\left(\sqrt{2E/m_x} - w\right)^2/u_0^2\right] - \exp\left[-\left(\sqrt{2E/m_x} + w\right)^2/u_0^2\right] \right\} \quad (2)$$

The resulting spectrum is shown n Figure 1 and has a peak near 0.3keV/GeV, so for a WIMP with a mass of 50 GeV, the peak in the spectrum is at 15 keV. Since the virial velocities are $O(10^{-3}c)$, typical scales for energies, momenta and masses are keV, MeV and GeV, respectively.

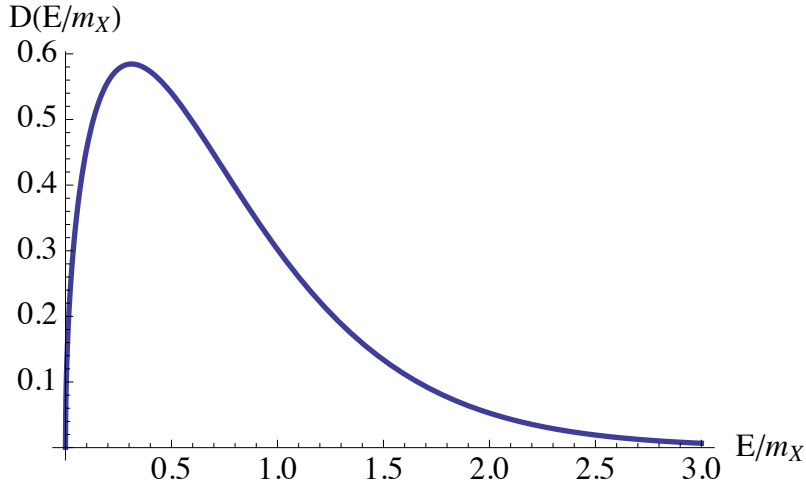

Figure1. WIMP spectrum in units of keV/GeV for typical values of w and $u_0$. The peak is in the neighborhood of 0.3 keV/GeV.

We now consider an elastic scattering of a WIMP with mass and momentum **p** from a nucleus with mass $m_N$ which recoils with momentum **q**. Integrating over all final states of the WIMP, the distribution of recoil momenta in the frame of the detector is

$$P(\mathbf{p},\mathbf{q}) = \frac{1}{4\pi}\left(\frac{M}{m_N p}\right)^2 \delta\left(\left|\mathbf{q} - \frac{m_N}{M}\mathbf{p}\right| - \frac{m_N}{M}p\right) \quad (3)$$

The average recoil energy for a WIMP of energy $E_X$ is

$$\langle E_R \rangle = \frac{2\mu}{M} E_X \quad (4)$$

where $\mu$ is the reduced mass. The average recoil energy is a maximum of $E_X/2$ for $m_X = m_N$.

The flux of WIMPs with velocity $v_X$ is $\Phi = n_X v_X = \rho_X p/m_X^2$ where $\rho_X \sim 0.3$ GeV/cm$^3$ is the local density of dark matter. The rate of recoils with momentum **q** is then

$$R(\mathbf{q}) = \frac{\rho_X A^2 \sigma_0 |S(q)|^2}{m_X} \int d^3v \, v \, f(\mathbf{v}) P(m_X \mathbf{v}, \mathbf{q}) \tag{5}$$

where $\sigma_0$ is the point-like one-nucleon cross section, $S(q)$ is the nuclear form factor normalized so that $S(0) = 1$, and $A$ is the atomic mass number. By (1) and (3) we find

$$R(\mathbf{q}) = \frac{R_0}{4\mu^2 w u_0 \pi^{3/2}} \frac{|S(q)|^2}{q} \exp\left[-\left(\frac{q/2\mu - \mathbf{w} \cdot \hat{\mathbf{q}}}{u_0}\right)^2\right] \tag{6}$$

where $R_0 = \rho_X \sigma_0 w / m_X$. The maximum of the exponential factor in (6) occurs for recoils parallel to the WIMP wind, $\mathbf{w}$, with momentum $q_{max} \sim 2\mu w$. Typical recoil momenta vary from $\sim 10$ MeV for very light WIMPs, e.g. $\mu \sim 5\text{GeV}$, to 200 MeV for heavy WIMPs, $m_X \sim m_N \sim 100$ GeV

A simple analytic form for the structure factor is due to Helm [28]

$$S(q) = \frac{3j_1(qr)}{qr} e^{-s^2 q^2 / 2} \tag{7}$$

where the parameters $R$ and $s$ are given by Lewin and Smith [1] as
$s = 0.9$ fm, $c = 1.23 A^{1/3} - 0.6$ fm, $a = 0.52$ fm, and $r = \sqrt{c^2 + \frac{7}{3}\pi^2 a^2 - 5s^2}$

The nuclear form factors for $^{127}$I and $^{27}$Na are shown in figure 2.

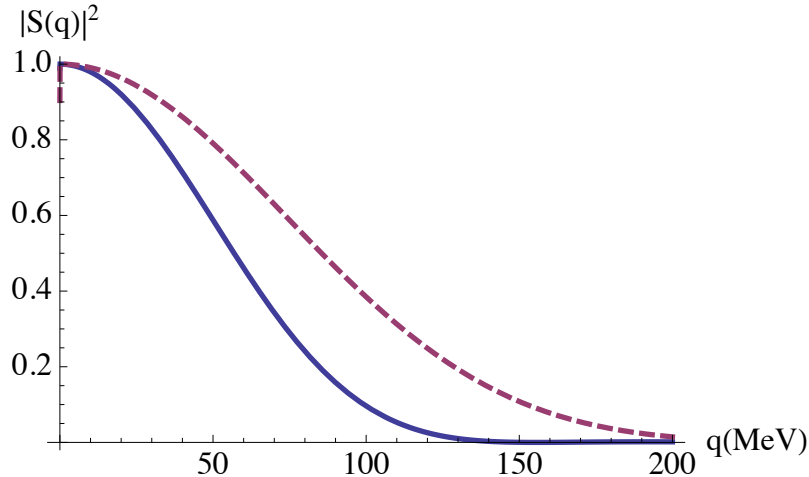

Figure2. The form factors for $^{127}$I (solid) and $^{23}$Na (dashed). Recoils with momenta greater than $\sim 200$ MeV are strongly suppressed.

From figure 2 we see that at momentum transfers typical of light WIMPs, the form factor can be ignored, while for heavier WIMPs it can reduce the cross section by a substantial factor. A detailed discussion of form factors was given by Helm and more recently by Duda, Kemper and Gondolo [21].

In order for directional variations to be significant, $R(\mathbf{q})$ must be peaked in the direction of he WIMP wind. The probability distribution for $x = \cos\theta$, where $\theta$ is the angle between the momentum of the recoiling nucleus, $\mathbf{q}$, and the WIMP wind, $\mathbf{w}$, for a given $q$ is

$$p(x;q) = \frac{2w}{\sqrt{\pi}u_0} \frac{\exp\left[-(q/2\mu - wx)^2/u_0^2\right]}{\mathrm{erf}\left[(q/2\mu - w)/u_0\right] - \mathrm{erf}\left[(q/2\mu + w)/u_0\right]} \tag{8}$$

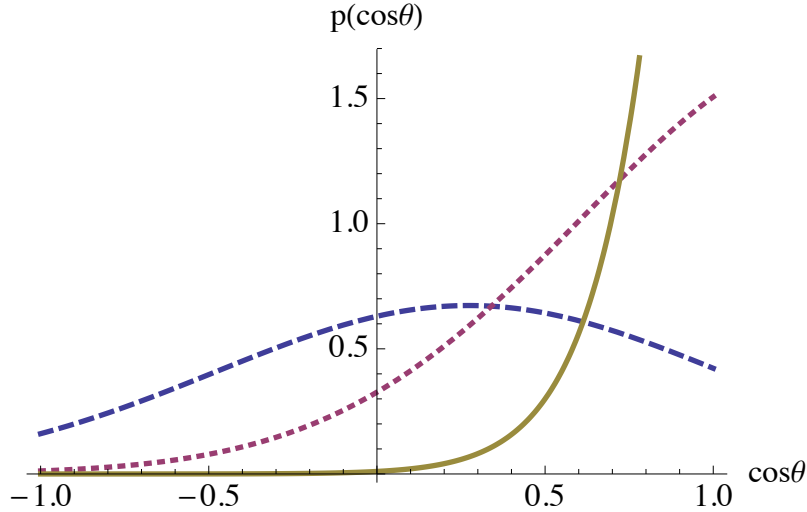

Figure 3. Angular distribution for elastic scattering for $q/\mu = 6.0\,\mathrm{MeV/GeV}$ (solid), $q/\mu = 2.0\,\mathrm{MeV/GeV}$ (dotted), and $q/\mu = 0.4\,\mathrm{MeV/GeV}$ (dashed). These distributions are normalized for different values of q, the momentum transfer. Note however that, as the normalization factor in the denominator of (8) indicates, the actual relative weight of different q's tends to strongly decrease with q.

As one can see from figure 3, the angular distribution at low $q/\mu$ is rather uniform, while at higher $q/\mu$ there is an enhancement in the direction of the WIMP wind. Roughly speaking, there is a significant fraction of recoils parallel to the WIMP wind if $q/\mu = 5.0\,\mathrm{MeV/GeV}$ or greater. In this case the energy of the WIMP, which must be greater than the recoil energy of the nucleus, satisfies the inequality $E_X/m_X > 12.5\,\mu/M$. If $m_X \sim m_N$ then, $E_X/m_X \sim 3.1$ well into the tail of the WIMP energy distribution shown in figure 1. Therefore come to the conclusion that kinematic considerations alone rule out directional effects if $m_X \sim m_N$. In order for the recoil momenta to correlate with the WIMP wind, $\mu/M \sim 0.04$, which means that for typical target nuclei, directional effects can be significant only for very light or very heavy WIMPs. Here we will consider the light-WIMP scenario.

Each type of detector has a threshold, typically a few keV, below which it is insensitive to recoils. A notable exception is the COGENT experiment [22] with a very low, $O(0.5\,\text{keV})$ threshold. This places a lower bound on the energy of a detectable WIMP, $E_X/m_X > E_{\min}/m_X$. For light WIMPs a threshold on the order of 2-3 keV or lower is needed; a higher threshold will again move the energy range of detectable WIMPs into the tail of the spectrum. Some types of detectors actually measure a fraction, Q (the quenching factor) of the recoil energy. If, as has been reported for NaI [23,24], the quenching factor is as small as 0.25, the threshold energy is increased by a factor of $1/Q$, making the detection of very light WIMPs impossible. However, the possibility that some recoil nuclei can "channel" between crystal planes with $Q \sim 1$ may make light WIMPs accessible. This will be discussed in detail in the next section.

The requirement that the recoil energy exceed the experimental cut-off excludes the majority of recoils at low momentum from being recorded. However, the events that do exceed the threshold and are recorded tend to have $q/\mu > 5.0\,\text{MeV/GeV}$. This is certainly the case for a 5 GeV WIMP scattering on iodine and to a lesser extent also for scattering on sodium with recoil energies in the 2.5-6 keV range where $q/\mu > 2.5 - 4.4\,\text{MeV/GeV}$. In this case a significant directional effect is possible.

### III. Channeling

The possible relevance of channeling to the DAMA-LIBRA experiment was recently emphasized by Drobyshevski [23]. He cited the earlier observations of Bredov and Okuneva [24] that 4keV Cs$^+$ ions can penetrate $\sim 10^3\,\text{Å}$ in germanium crystals versus $44\,\text{Å}$ in amorphous germanium. Channeling can significantly enhance the "quenching factor", that is the ratio of electromagnetic energy deposited by a recoiling nucleus to that of a photon of the same energy. The quenching factor in NaI(Tl) has been measured several times [25] with values in the range 0.19-0.26.

The calculation presented by Bernabei et al [5] suggests that for Na or I ions recoiling along a channel, the quenching factor may be close to one. If this is correct, then channeling, which is more prevalent at low energies, generates events with higher scintillation outputs. Further, if WIMPS are light, on the order of 5 GeV, then *only* channeled recoil ions will deposit enough energy to be detected.

Given the recoil momentum, **q**, the energy detected as scintillation light is

$$E = \frac{q^2}{2m_N} Q(\mathbf{q}) \quad (9)$$

where $Q$ is the quenching factor (this is a slightly non-standard definition.) A recoil ion can channel between two atomic planes if the transverse momentum is below a cutoff, $q < k_0$, which means the acceptance angle for channeling scales as $1/\sqrt{E}$. Typical acceptance angles for 4 keV $^{127}$I ions channeling in NaI ions are in the range of a few degrees.

A simple model of the probability of channeling is the bimodal distribution

$$P(Q,\mathbf{q}) = (1 - p_C(\mathbf{q}))\delta(Q - Q_0) + p_C(\mathbf{q})\delta(Q - Q_1) \quad (10)$$

where $p_C(\mathbf{q})$ is the probability that an ion with momentum $\mathbf{q}$ channels. Figure 4 shows the probability of channeling for a single family of atomic planes, as a function of the angle between the recoiling ion and the normal to the planes. Here we use a simple empirical model for the probability to channel,

$$p_j(\mathbf{q}) = p_B e^{-(\hat{\mathbf{n}}_j \cdot \mathbf{q})^2 / k_0^2} \quad (11)$$

where $\hat{\mathbf{n}}_j$ is the normal to a given family of planes, $k_0$ is the momentum cutoff that determines the acceptance angle for channeling, and $p_B$ is the probability that the ion is not blocked. Typical acceptance angles for channeling are on the order of 0.1 rad. For a recoiling iodine ion with a momentum of 30 MeV (4 keV kinetic energy), we take $k_0 \sim 3$ MeV. Figure 4 shows the probability to channel as a function of the angle for 30 MeV recoils with $p_B = 0.25$ for a single atomic plane.

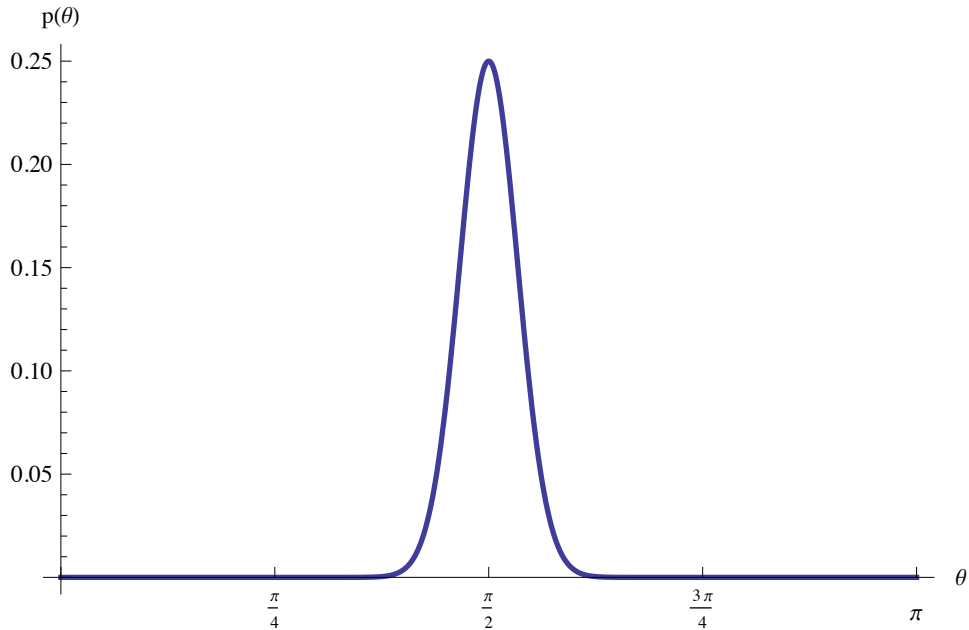

Figure 4. Probability to channel as a function of the angle between he recoil momentum and the normal to a single family of planes for q=30 MeV, $k_0 = 3\text{MeV}$ and $p_B = 0.25$.

If the ion does not channel, a fraction of the recoil energy, $Q_0$, is deposited in electromagnetic degrees of freedom. If the ion does channel, then a fraction $Q_1 > Q_0$ of the energy is deposited. In a highly symmetric crystal like NaI there can be several possible families of planes along which an ion with momentum **q** can channel. If we treat channeling between different planes as statistically independent events, the channeling probability is

$$p_C(\mathbf{q}) = \sum_{k=1}^{N} p_k(\mathbf{q}) y_k(\mathbf{q}) \tag{12}$$

where

$$y_k(\mathbf{q}) = \prod_{j=1}^{k-1} \left[1 - p_j(\mathbf{q})\right] \tag{13}$$

If we take the rate of recoils with momentum **q**, (6) and fold it with the distribution of quenching factor, (10), the measured energy spectrum is

$$I(E, \mathbf{w}) = I_0(q_0)\left[1 - f(q_0, \mathbf{w})\right] + I_0(q_1) f(q_1, \mathbf{w}) \tag{14}$$

where

$$I_0(q_0) = \frac{R_0 m_N}{4\mu^2 w^2} \frac{|S(q_0)|^2}{Q_0} \left[\text{erf}\left(\frac{q_0/2\mu + w}{u_0}\right) - \text{erf}\left(\frac{q_0/2\mu - w}{u_0}\right)\right] \tag{15}$$

is the rate of detection in the absence of channeling, $q_0 = \sqrt{2m_N E / Q_0}$,

$$f(q, \mathbf{w}) = \frac{\int d^2\hat{q} \exp\left[-\left(\frac{q}{2\mu u_0} - \frac{\hat{\mathbf{q}} \cdot \mathbf{w}}{u_0}\right)^2\right] p_C(\mathbf{q})}{\int d^2\hat{q} \exp\left[-\left(\frac{q}{2\mu u_0} - \frac{\hat{\mathbf{q}} \cdot \mathbf{w}}{u_0}\right)^2\right]} \tag{16}$$

is the probability to channel averaged over the recoil distribution for a given momentum transfer and average WIMP velocity.

and $q_1 = \sqrt{2m_N E/Q_1}$. If there is a diurnal modulation due to channeling, then it will show up through the time-dependence of $f(q,\mathbf{w})$ as the orientation of the detector varies with respect to the WIMP wind.

In figure 5 we show the function $f(q,\mathbf{w})$ for q=30MeV over a single day. The exact time dependence varies from day to day but the variation from maximum to minimum is always on the order of 0.1%. Note that while channeling does not give rise to a large diurnal variation, it does significantly increase the effective quenching factor.

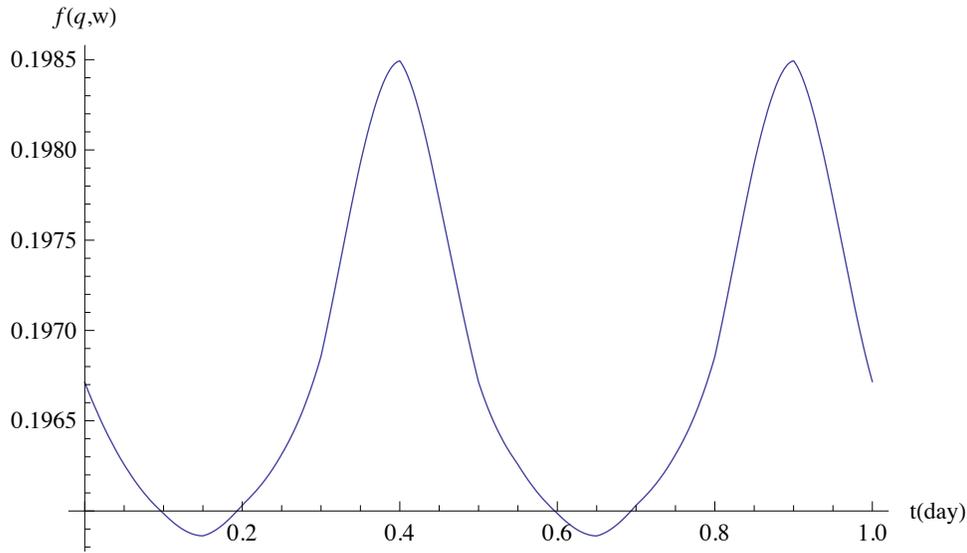

Figure 5. Probability of channeling averaged over the recoil distribution for q=30MeV (3.8 keV), $m_X$=5GeV, $p_B = 0.25$ and w=250 km/s.

While the above model can give some general idea as to the importance of channeling, clearly an experimental study extending the neutron scattering studies of Chagaini et al [25] is needed. In the latter experiment a monochromatic beam of *one* specific energy was used and the recoil energy inferred from the angle by which the neutron scattered. Having such measurement at *several* well defined neutron energies in the range of a few MeV would help disentangle the effect of changing the magnitude of the recoil momentum and its direction, allowing us to measure the relevant input for our analysis - the recoil direction dependence of the scintillation response for each recoil energy separately.

## IV. Discussion

Most WIMP research is focused on "symmetric" models where the relic WIMP density, $\rho_X$, freezes out with $X - \bar{X}$ annihilations. SUSY models can provide $m_X \sim O(100 \text{ GeV})$ stable LSP (lightest SUSY partner) whose weak scale annihilation cross-section yields the required $\rho_X$. The fact that $m_X$ roughly matches the masses of the Ge, I and Xe nuclei used in most detectors tends to maximize the WIMP-nucleus cross section and recoil energies, helping direct WIMP searches. Also, present-day annihilations generate an array of possible indirect WIMP searches.

Light WIMPs, $m_X \sim O(10 \text{ GeV})$, were suggested in order to explain the DAMA-LIBRA annual modulation and more recently in connection with COGENT and CDMS results[26,27]. Light WIMPs arise naturally in asymmetric models [28-32] where only the excess $\Delta n(X) = n(X) - n(\overline{X})$ remains after efficient $X - \overline{X}$ annihilations. If we assume that the X-asymmetry is similar to the baryon asymmetry, then, since $\Omega_X \sim 6\Omega_B$, $m_X \sim 5 - 6 \text{ GeV}$. In these models indirect WIMP searches based on $X - \overline{X}$ annihilations are no longer possible. Also, direct searches seem to be more difficult: For a given X-nucleon cross-section, the cross section for a nucleus with atomic number, A, is smaller and most importantly the recoil energies, especially for $m_N \gg m_X$, are now much reduced. However, as we have seen above, it is precisely in such cases that requiring some minimal observable recoil dramatically increases the alignment of the WIMP wind and nuclear recoil, and revives the possibility of observing the diurnal modulation.

Regardless of how likely such a scenario is, it is natural to simply take the events observed in single crystal detectors and plot them vs. time modulo the sidereal day. No background of either terrestrial or solar origin can yield a modulation that is periodic in the sidereal day over ten years. Further, the phase of the modulation is fixed by the same WIMP wind direction that controls the phase of the annual modulation. Further, the pattern of variation in a sidereal day should reflect the symmetry of the crystal in the detector and the latitude of the experiment. Hence, finding such modulations even at relatively low levels could provide crucial corroborating evidence for WIMPs . Our above discussion indicates however that the reverse is not true. For heavy WIMPs crystalline detectors – even with significant over-all channeling – are unlikely to exhibit diurnal modulation, except in the high-energy tail of the recoil spectrum.

Acknowledgements: This work was partially supported by the National Science Foundation Grants PHY-0500337 and PHY 0855314. We would also like to thank Graciela Gelmini for helpful discussions, and for forwarding a draft of the article on Channeling in Direct Dark Matter Detection.